\documentclass[conference]{IEEEtran}

\IEEEoverridecommandlockouts

\usepackage{cite}
\usepackage{comment}
\usepackage{amsmath,amssymb,amsfonts} 
\usepackage{algpseudocode}
\usepackage[linesnumbered,ruled]{algorithm2e}
\usepackage{algpseudocode}
\usepackage{graphicx} 
\usepackage{float} 
\usepackage{textcomp} 
\usepackage{eso-pic}
\usepackage{textcomp} 
\usepackage{makecell}
\usepackage{multicol}
\usepackage{multirow}
\usepackage{caption}
\usepackage{tabularx,booktabs}
\usepackage{xcolor}
\usepackage{amsmath,amssymb,amsfonts} 
\usepackage{algpseudocode}
\usepackage[linesnumbered,ruled]{algorithm2e}
\usepackage{algpseudocode}
\usepackage{graphicx} 
\usepackage{float} 
\usepackage{textcomp} 
\usepackage{eso-pic}
\usepackage{xcolor}
\usepackage{makecell}
\usepackage{multicol}
\usepackage{multirow}
\usepackage{caption}
\usepackage{tabularx,booktabs}
\newcolumntype{C}{>{\centering\arraybackslash}X} 
\usepackage{array}
\newcolumntype{P}[1]{>{\centering\arraybackslash}p{#1}}
\usepackage{comment}
\usepackage{csquotes}
\usepackage{lineno,hyperref}

\usepackage{csquotes}
\def\BibTeX{{\rm B\kern-.05em{\sc i\kern-.025em b}\kern-.08em
    T\kern-.1667em\lower.7ex\hbox{E}\kern-.125emX}}
    
\makeatletter
\def\ps@IEEEtitlepagestyle{%
  \def\@oddfoot{\mycopyrightnotice}%
  \def\@evenfoot{}%
}
\def\mycopyrightnotice{%
 {\footnotesize 978-1-7281-6309-3/19/\$31.00 \textcopyright2019 IEEE\hfill}
  \gdef\mycopyrightnotice{}
}

\usepackage[textsize=footnotesize, textwidth=1.2cm]{todonotes}

\usepackage{eso-pic}

\specialcomment{fpal}{\color{red}}{\color{black}}

\begin{document}




\title{Interoperability and Explicable AI-based Zero-Day Attacks Detection Process in Smart Community}

\author{\IEEEauthorblockN{Mohammad Sayduzzaman\IEEEauthorrefmark{1},  Anichur Rahman\IEEEauthorrefmark{2}, Jarin Tasnim Tamanna\IEEEauthorrefmark{3},  Dipanjali Kundu\IEEEauthorrefmark{4}, and Tawhidur Rahman \IEEEauthorrefmark{5}}
\IEEEauthorblockA{\textit{Department of Computer Science and Engineering}, \\
\textit{National Institute of Textile Engineering and Research (NITER)}\\
\textit{Senior Specialist, Digital Security \& Digital Diplomacy, ICT Division, Agargaon, Dhaka-1207} \\
msayduzzaman@niter.edu.bd\IEEEauthorrefmark{1},
anis\_cse@niter.edu.bd\IEEEauthorrefmark{2},
jtasnim@niter.edu.bd\IEEEauthorrefmark{3},
dkundu@niter.edu.bd\IEEEauthorrefmark{4} and pialfg@gmail.com\IEEEauthorrefmark{5}}}

\maketitle

\begin{abstract}
\boldmath
Systems, technologies, protocols, and infrastructures all face interoperability challenges. It is among the most crucial parameters to give real-world effectiveness. Organizations that achieve interoperability will be able to identify, prevent, and provide appropriate protection on an international scale, which can be relied upon. This paper aims to explain how future technologies such as 6G mobile communication, Internet of Everything (IoE), Artificial Intelligence (AI), and Smart Contract embedded WPA3 protocol-based WiFi-8 can work together to prevent known attack vectors and provide protection against zero-day attacks, thus offering intelligent solutions for smart cities. The phrase \enquote{zero-day} refers to an attack that occurs on the \enquote{day zero} of the vulnerability's disclosure to the public or vendor. Existing systems require an extra layer of security. In the security world, interoperability enables disparate security solutions and systems to collaborate seamlessly. AI improves cybersecurity by enabling improved capabilities for detecting, responding, and preventing zero-day attacks. When interoperability and Explainable Artificial Intelligence (XAI) are integrated into cybersecurity, they form a strong protection against zero-day assaults. Additionally, we evaluate a couple of parameters based on the accuracy and time required for efficiently analyzing attack patterns and anomalies. 
\end{abstract}
\vspace{2mm}

\begin{IEEEkeywords}
Zero-day Attack, Intrusion Detection System, Artificial Intelligence, Internet of Everything, Explainable AI, Cybersecurity, Interoperability, Data Analysis.
\end{IEEEkeywords}

\section{Introduction}
Attackers frequently use zero-day exploits to obtain unauthorized system access, steal sensitive information, disrupt services, or execute malicious code without being detected. As there is no prior information or defense against zero-day vulnerabilities, these attacks can be extremely harmful and difficult to remediate. Existing signature-based detection systems prove inefficient against zero-day attacks due to the Lack of Signatures, no prior knowledge, and, most importantly, polymorphic characteristics of malware \cite{kumar2024detection, Rahman2020}. While some attackers employ methods like polymorphism, which alters virus properties over time, signature-based systems find it challenging to detect any anomalies or attacks. The anomaly-based detection system also failed to detect the newest attack as it lacks historical data as well as a sensitivity vs. false positive attitude. Attackers may employ sophisticated evasion techniques or mimic legitimate traffic to avoid triggering alerts and, most importantly, limited scope, which may not be adequately covered by the conventional detection system \cite{ahmad2023zero}. AI algorithms can effortlessly combine and analyze network logs, user activity analytics, endpoint security tools, and other data sources. This comprehensive approach improves threat detection accuracy. AI can swiftly assess data from interoperable systems and automate the reaction to zero-day assaults, minimizing the time window for an attacker to exploit the vulnerability. Interoperability enables companies and security systems to work together and share threat intelligence. In this research, we present an approach that combines interoperability with explainable artificial intelligence (XAI) to discover an optimal solution for detecting zero-day attacks \cite{rizzardi2024nero, rahman2021study}. \vspace{2mm}

\begin{figure}[h]
    \centering
    \includegraphics[scale=0.28]{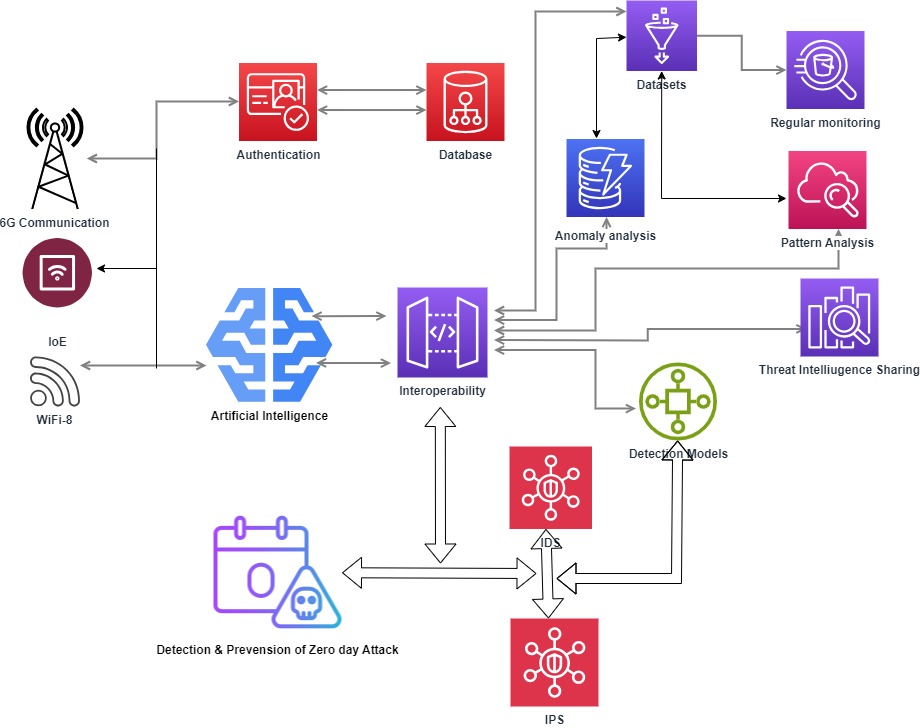}
   \caption{\textcolor{black}{Concept of Zero-day attack detection through interoperability, and AI}}
    \label{fig:f1}
\end{figure}

Interoperability facilitates seamless communication between disparate systems. This is necessary to combine different technologies into a coherent network, including Wi-Fi 8, 6G, and the Internet of Everything IoE \cite{rahman2024machine, rahman2024internet}. Fig. \ref{fig:f1} explains the concept of interoperability in the context of technology, which refers to making sure that various platforms, networks, and devices can communicate with each other and function as a single unit, particularly in the areas of networking and cybersecurity. Systems can cooperate regardless of the underlying architecture when they are constructively interoperable \cite{rahman2023impacts}. Creating integrated networks—especially helpful for IoE ecosystems and multi-network environments—where different devices and technologies coexist and collaborate requires cross-platform interoperability. Through the facilitation of unified threat identification and response, interoperability can enhance security. Faster incident response and a wider understanding of security threats are made possible by this shared knowledge. Information sharing between systems makes it simpler to identify and address security risks, such as zero-day attacks. Interoperability guarantees that new technologies can be integrated into current systems with minimal effort\cite{jinhong2024cross}.

\vspace{2mm}
A cyberattack that leverages an undiscovered software vulnerability is known as a zero-day assault. Because the vulnerability is unknown to the software vendor or security community, no patch or remedy was available at the time of the attack. Fig. \ref{fig:f2} describes the life cycle of a zero-day attack, highlighting its essential stages \cite{9528133}:

\begin{figure}[h]
    \centering
    \includegraphics[scale=0.33]{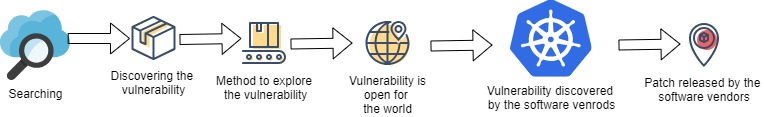}
   \caption{\textcolor{black}{Life Cycle of a Zero Day Attack}}
    \label{fig:f2}
\end{figure}
\begin{itemize}
    \item \textbf{Search and Discover}: A zero-day vulnerability is found by an individual or group, such as a researcher, hacker, or member of a security organization. The discovery could be made by code analysis, fuzzing, or exploiting other security issues.\vspace{2mm}
    
    \item \textbf {Development and Weaponizing}: Once the vulnerability has been identified, attackers create an exploit to take advantage of it. This entails writing code or developing a method for exploiting the vulnerability to infiltrate systems. In some circumstances, attackers may tailor the exploit to specific targets or conditions. \vspace{2mm}
    
    \item \textbf{Deploy and Explore}: The exploit is converted into a format that can be utilized in a real-world attack. This could include generating malware, incorporating the exploit into existing malware, or launching a phishing effort to deliver the exploit to targets. Attackers frequently package their exploits to avoid detection by security software.\vspace{2mm}
    
    \item \textbf{Impact and Disclosure}: Exploitation of the zero-day vulnerability can have serious implications. This could include data breaches, ransomware attacks, financial losses, espionage, and other harmful acts. The impact is often determined by the importance of the target, the sort of data or systems compromised, and the attacker's goals.\vspace{2mm}
    
    \item \textbf{Response, Mitigation, and Learning}: The software provider provides a patch or upgrade to address the vulnerability. This approach can take some time, particularly if the vulnerability is complicated or impacts key systems. Security teams may undertake postmortem investigations to determine how the zero-day attack occurred, assess its tactics, and identify security process improvements to prevent such attacks in the future.
    
\end{itemize}

Though this is a well-renowned life cycle, an attacker might change or skip any steps mentioned here. An Intrusion Detection and Prevention System (IDPS) operates by keeping an eye on system activity or network traffic, analyzing collected data for threats using a variety of techniques, identifying potentially malicious traffic, notifying relevant personnel of identified threats, taking appropriate action to stop intrusions once they are discovered, and producing reports for analysis and compliance \cite{rahman2022integration}. 
IDPS faces significant challenges in detecting zero-day attacks, as zero-day take advantage of undiscovered vulnerabilities, which prevents IDPS from having the required signatures or patches. Zero-day attacks do not have the known threat signatures that traditional signature-based IDPS rely on, making these systems unreliable against them. In our proposed methodology to overcome this issue, we introduced an intermediate layer between interoperability and the IDPS system that added an extra layer of defense through XAI to detect the zero-day attack. The main contribution of the paper is--
\begin{itemize}
    \item To boost IDPS's overall performance, this work couples Artificial Intelligence with interoperability among sixth-generation (6G), WiFi-8, and IoE.
     \vspace{2mm}
     
    \item The authors focus on Zero-Day attack pattern detection by combining Machine Learning with Explainable AI (XAI). \vspace{2mm}
    
    \item In addition, they evaluate multiple models and consolidate the results for further investigation.
\end{itemize}
\vspace{2mm}

\textcolor{black}{\textbf{Organization:} This paper is structured as follows: Discussions on the previous research and the concept of Interoperability among 6G, IoE \& Wifi-8 in section II, Zero-day attack and its lifecycle to prevent the attack finally after weaponizing phase. Then, the proposed AI-based IDPS for zero-day attack detection, along with prevention and procedures, is presented in section III. The dataset description part is presented in section IV. Moreover, results and a related discussion are given in section V. Lastly, section VI contains the conclusion and some thoughts, limitations, and future scopes. }

\section{\textcolor{black}{Related Works}}
Intrusion Detection Systems have been essential in identifying any abnormal or suspicious activity that could compromise general security and pave the way for significant cyberattacks since the invention of Wi-Fi technology \cite{32}. Many researchers employ various techniques to identify and stop zero-day attacks \cite{31}. A few work with heuristic analysis, others with behavioral analysis, others with signature-based detection, others with threat intelligence, others with endpoint protection, others with advanced threat detection solutions, and others with cloud-based solutions \cite{33}. \vspace{2mm}

Kumar et al. \cite{kumar2021robust} study discussed parameters that are overlooked when identifying zero-day attacks. By omitting those requirements, several organizations claim they can manage complex cyberattacks but not in practical scenarios. By ignoring those requirements, we are merely avoiding the risk, and the attacker manages to complete his mission. Hindy et al. \cite{hindy2020utilising} use a deep learning-based methodology to identify zero-day attacks, with the primary risk being that the decision-making process frequently misses or fails to detect the true attack. Similarly, M. Macas et al. \cite{34} focused on different deep learning technologies for detecting attacks in various perspectives.\vspace{2mm}

Zhang et al. \cite{zhang2021novel} showed how aggregated vulnerability-based assessment could detect zero-day attacks. Martins et al. \cite{martins2022host} worked on a host-based detection system. Another paper Salim et al. \cite{salimarticlesfederated} implemented federated learning based detection system for healthcare. Zahoora et al. \cite{zahoora2022ransomware} mitigated XML injection based zero day attack via strategy-based detection system. On the other hand, Efe et al.  \cite{efe2022comparison} shows the comparison of host-based and network-based detection systems and what their limitations are. IDPS may be anomalous or signature-based as Nie et al. \cite{35} focused on existing strategies and their loopholes. Recently, they began combining different technologies, such as SC and ML, to boost performance. We will describe the adoption process for AI and interoperability, followed by a review of previous work.\vspace{2mm}

From the above discussion, we summarise that Numerous publications on Intrusion Detection Systems (IDS) and Intrusion Prevention Systems (IPS) based on ML \& AI were discovered throughout our investigation on this subject\cite{talukder2023dependable}. In certain articles, AI and ML are combined for IDS or IPS. However, there aren't many papers that show how interoperability among cutting-edge state-of-the-art technology with AI provides threat detection feeds for IDPS that prevent zero-day attacks \cite{36}. That is the reason for our actions. For intrusion detection, we address XAI along with recent technologies. 

\section{Proposed Methodology for Interoperability and Explainable Artificial Intelligence (XAI)-based IDPS}

\begin{figure*}[!ht]
    \centering
    \includegraphics[scale=0.43]{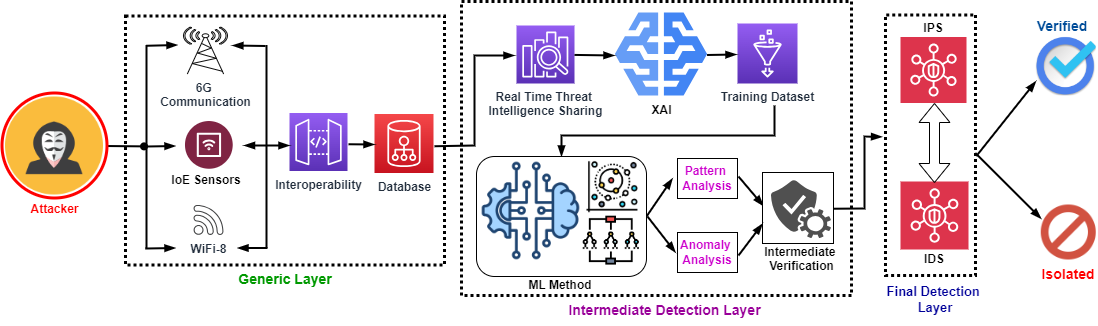}
   \caption{\textcolor{black}{Proposed Architecture for Interoperability and XAI-based IDPS}}
    \label{fig:f3}
\end{figure*}

In the constantly changing field of cyber security, one of the most dangerous risks that businesses and individuals encounter is the feared zero-day attack.  The vendors are left with little time to create and implement a patch since these attacks make use of flaws in hardware, software, or firmware that they are unaware of \cite{rahman2020distblockbuilding, rahman2020distb}.  Hence, until a fix is released, hackers may take advantage of these vulnerabilities to obtain unauthorized access, interfere with business processes, or steal sensitive data. To overcome this issue, we developed an explainable artificial intelligence (XAI) based zero-day attack detection system given in Fig. \ref{fig:f3}. Most importantly, the proposed methodology is divided into three sections: the generic layer, the intermediate layer, and the final detection layer. In the generic layer, real-time thread intelligence of smart community systems (Wi-Fi8, IoE, 6G communication) is shared by interoperability to the intermediate layer \cite{10103295}. In the intermediate layer, we made use of machine learning techniques with XAI to create a robust zero-day attack detection system. The final layer analyzes the threat level and sends alerts to security teams through the IDPS system efficiently \cite{rahman2021smartblock, 9499121}.

\subsection{\textcolor{black}{Generic Layer Discussion with Interoperability and Zero-Day Attacks}}
In case of cybersecurity, interoperability is the capacity of various hardware, software, and application platforms to efficiently exchange data, collaborate, and communicate with one another. It shares real-time threat intelligence, providing a more robust and timely response to emerging threats that can lead to faster identification of potential zero-day attacks. Interoperable security solutions allow for the correlation of events between various systems and network segments \cite{rahman2023icn}. This connection can assist in identifying trends or abnormalities that might point to the existence of a zero-day attack that one system might not be able to detect. For example, if IoE detects suspicious activity that could be a zero-day attack, it can inform other systems like 6G communication and Wi-Fi8 to take preventative action without human intervention. Users are always creating data,  a large amount of data is generated per day which is about 100 zettabytes \cite{edvardsson1999survey, rahman2022sdn} by Wi-fi8, IoE sensors, and 6G communication systems. To manage this large data most efficiently and securely in the generic layer, we used interoperability, which connects with the intermediate layer through real-time thread sharing. 

\subsection{\textcolor{black}{Intermediate Layer Detection through ML and XAI Training Process}}
As previously described, zero-day attacks are mostly unpredictable by conventional ML and IDPS techniques, as their pattern is unrecognizable. To address this issue, in our proposed method, we applied explainable AI (XAI), which enables human users to transparently and easily comprehend the decisions and behaviors of AI systems with a high degree of learning performance (accuracy). From the generic layer, we will get the real-time sharing threat (data) and create a training database for generating shape value via XAI. In machine learning, SHAP (SHapley Additive exPlanations) values are a technique for analyzing any model's prediction. 

\begin{figure}[h]
    \centering
    \includegraphics[scale=0.50]{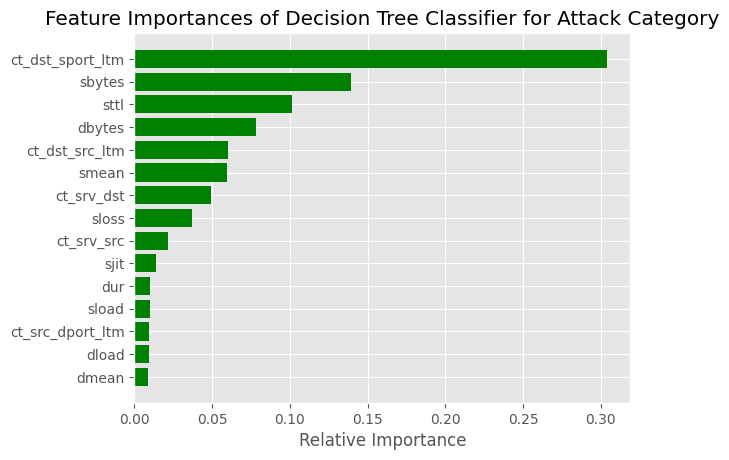}
   \caption{\textcolor{black}{SHAP values for Attack Pattern Analysis}}
    \label{fig:f4}
\end{figure}

\begin{figure}[h]
    \centering
    \includegraphics[scale=0.50]{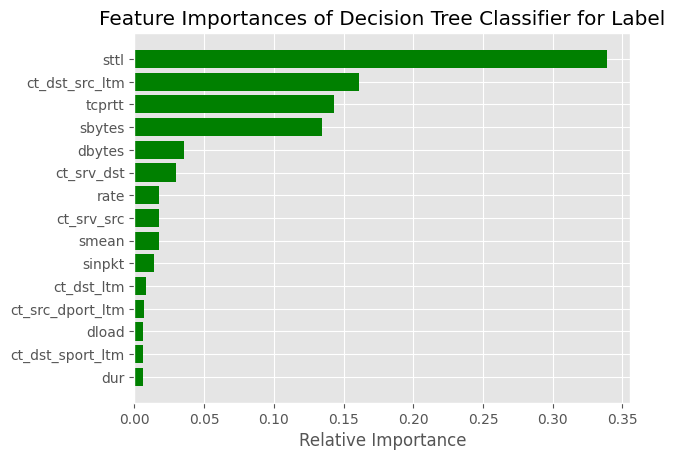}
   \caption{\textcolor{black}{SHAP values for Anomaly Analysis}}
    \label{fig:f5}
\end{figure}

\vspace{2mm}
In this paperwork, we apply XAI techniques on the dataset given in to generate two SHAP values, one for anomaly analysis and another for attack pattern analysis, which is the most significant feature for zero-day attack detection. In this dataset, there are nearly 45 attributes that are responsible for anomaly detection and attack pattern analysis. After applying XAI, we get 15 optimal attributes given in \ref{fig:f4} and \ref{fig:f5}, indicating the contribution of each feature that is responsible for the attack and its pattern. In both figures, the Y-axis indicates the list of features that are highly responsible for the attack and its pattern detection, while the X-axis represents the severity index of the list of features. Both the SHAP values are generated based on the equation \ref{eq1} given below \cite{moustafa2016evaluation, udoy20234sqr} :

\begin{equation}
\label{eq1}
f(x) = \sum_{i=1}^{P} \phi_i + E_x[f(x)]
\end{equation}

Where the model has $p$ features, and $\phi_i$ is the SHAP value for feature $i$. According to this equation, the average prediction and the total of all SHAP values equal the prediction for that particular occurrence.\vspace{2mm}

Fig. \ref{fig:f4} shows the SHAP values that are responsible for attack pattern detection, which is the main contribution of our work. As we know, zero-day can not be identified because it has no known pattern. However, XAI can detect the features that are responsible for the attack pattern and detect newly encountered attack patterns very efficiently. On the other hand, Fig. \ref{fig:f5} shows the SHAP values that are responsible for anomaly detection, which is also possible in conventional IDPS systems. However, after applying XAI the required computational time is decreased, on the contrary accuracy increased. This approach works as a two-layered detection system, the known attack patterns which already available are detected in this layer by the ML approach. As the zero-day is not possible to detect in this manner, it is passed to the final detection layer. As newly encountered zero-day attack patterns are identified via XAI in the intermediate layer and passed to the final detection layer, they can easily be identified via the IDPS technique \cite{ahmed2023image}.

The intermediate layer is the main contribution of this paperwork, which resolves the issues of the IDPS system in case of detection of a zero-day attack. As the pattern of zero-day is unknown, it can not be resolved by a conventional IDPS system, that's why we introduced an intermediate layer where unknown zero-day attack patterns are detected in an efficient manner and passed to the IDPS to detect in the final layer detection system. Step by step, the working procedure for preventing zero-day attacks is mentioned as follows--
\begin{itemize}
    \item \textbf{Step 1 (Infrastructure Setup):} Create a network structure that combines all 6G connections, IoE devices, and Wi-Fi 8 access points. Configure a multi-layered security system that includes firewalls, IDS/IPS, and endpoint protection.\vspace{2mm}

    \item \textbf{Step 2 (Data collection):} Establish real-time thread sharing through interoperability for all communication channels on 6G, IoE, and Wi-Fi 8 networks. \vspace{2mm}

    \item \textbf{Step 3 (Analysis):} Use XAI with ML technologies to analyze network traffic and detect anomalies in the intermediate layer. A Security Operations Center (SOC) or equivalent system should be used to monitor the network at all times \cite{demertzis2019next}. All critical security-related incidents should be documented for further analysis and auditing. \vspace{2mm}

    \item \textbf{Step 4 (Detection and Alert):} Monitor any unusual activities through IDPS techniques in the last layer then compare and check it with the treat intelligence stream. Provide security alerts using SOC if any anomaly occurs in the communication system.\vspace{2mm}

    \item \textbf{Step 5 (Recovery):} Maintain regular backups of essential data and develop recovery plans. Conduct a thorough analysis to determine the underlying cause of a zero-day attack. Apply the appropriate updates and remedial measures to address vulnerabilities.   
\end{itemize}

\subsection{\textcolor{black}{Final Layer Anomaly Detection with IDPS Technique}}
The types of attacks that are undetectable in the intermediate layer, like zero-day, will be detected in the final detection layer. As mentioned earlier, zero-day can not be detected by conventional IDPS systems, so we introduced an intermediate layer between interoperability and IDPS that will detect the newly encountered zero-day attack pattern using XAI techniques. This pattern will be passed to the IDPS system and it will block the user if there is any anomaly, otherwise pass the user normally. One of the most commonly used metrics is utilized in this work to assess the efficiency of intrusion detection models accuracy \cite{soltani2023adaptable, rahman2023towards}.

\section{Datasets Description}
This section focused on the proposed framework from a variety of perspectives and datasets. Initially, cupKDD99 was used to detect an intrusion. The effect of detection on performance is examined. The parameters with the best results are then picked for comparison with the centralized approach. The SHAP results are then used to explain and understand the outcomes of the proposed framework. The study concludes with a thorough discussion of the findings, providing insightful interpretations. NSL-KDD datasets are used to determine the categories of different types of attacks \cite{dhanabal2015study, khan2022multinet}. The NSL-KDD dataset is commonly utilized in intrusion detection and cybersecurity research. It is an improved version of the original KDD Cup 1999 dataset, which addresses some of its shortcomings. The KDD Cup 1999 dataset was built on a simulation of a military network and had various flaws, including excessive redundancy and poor representation of contemporary network traffic. The NSL-KDD dataset overcomes these shortcomings by presenting a more balanced and realistic depiction of network traffic \cite{kundu2024federated}. It is frequently used to assess intrusion detection systems (IDS) and machine learning models in the context of cybersecurity. Despite its improvements over the original KDD Cup 1999 dataset, NSL-KDD still contains certain flaws, such as redundant features and a lack of representation for some attacks \cite{Islam2021, Debnath2022}.\vspace{2mm}  

UNSW-NB15 is used for network traffic feature extraction \cite{moustafa2016evaluation}. The UNSW-NB15 dataset is another extensively used dataset in network security and intrusion detection. It was developed by academics at the University of New South Wales (UNSW) in Australia and is specifically intended for evaluating network intrusion detection systems (NIDS). Another type of intrusion detection is host-based detection systems (HBDS). The dataset is based on network traffic recorded in a controlled setting, which may not accurately reflect the complexity and variety of real-world network traffic. As a result, models trained on this dataset may not perform well in real-world circumstances. Though it has some limitations, it is very helpful for security researchers. We use this dataset to check if our system is able to detect any unknown attacks. The result is satisfactory as the system can successfully detect unknown attacks and prevent them by successfully isolating from the entire network through IDPS. The ToN-IoT dataset is a useful resource for evaluating the effectiveness of AI-enabled cybersecurity applications across IoT, network traffic, and operating systems \cite{moustafa2021new}. The ToN-IoT dataset, like other intrusion detection datasets, may suffer from class imbalance, in which the number of cases reflecting typical traffic outnumbers those showing malicious or aberrant behavior. This mismatch can have an impact on the performance of machine learning models as well as how assessment results are interpreted \cite{rahman2024blocksd}.

\section{Result Analysis and Performance Measurement}

In Table \ref{Tab1}, we compared our proposed work with some of the existing work considered during the background study. Moreover, for the experimental analysis in this work, the following parameters are used while measuring the performance and analyzing the results.
\begin{itemize}
    \item Security Metrics: Security metrics include zero-day attack detection rate, false positive rate, incident response time, attack surface reduction, and anomaly detection accuracy.  The accuracy is calculated by using Equation 2.
    \begin{align} Accuracy=&\frac {TP+TN}{TP+TN+FP+FN},  \end{align}

    \item Efficiency Metrics: It focuses on resource consumption, network latency, algorithm processing time, and scalability.

    \item Effectiveness: It includes Threat prevention rate, attack vector identification, Incident resolution rate, and defensive capability.

    \item Continuous Improvement: We prioritize feedback loop efficacy, patch management efficiency, and security training impact for optimal performance. 
\end{itemize}

\begin{table*}[]

\centering
 \caption{\textcolor{black}{Comparison of the proposed framework with the other state of art models}}
\label{tab:comp}
\begin{tabular}{ |p{2.5cm}|P{2.5cm}| P{3.5cm}| P{2.2cm}| P{3.5cm}|}
    \toprule
    Method & {Number  of Attributes} & Classifier & {Classification Type} &{Accuracy of  Attack Detection} \\
    \hline

     Sarhan et al. \cite{sarhan2023zero} & 24,3064 & MLP and RF & 0ne Class & 85.5\% \\
    \hline
  
    Hindy et al. \cite{hindy2020utilising} & 19,663 & SVM and Autoencoders & One Class & 92.96\% \\ 
    \hline

    Kumar et al. \cite{kumar2021robust} & 53,234 & Hitter and Graph Technique & Multi Class &  88.98\%  \\
    \hline
   Koroniotis et al.  \cite{koroniotis2020new} & 18,563 & Decision Tree & Binary & 93.2\% \\
   \hline

    Zahoora et al. \cite{zahoora2022ransomware} & 16382 &   CSPE-R Ensemble  & Binary & 93\%\\
    \hline
   Ashraf et al.  \cite{ashraf2020novel} & 18,563 & FSVM  & Multi-Class & 92\%\\
   \hline
    
    Proposed Method	& 18,563 & XAI and ML & Multi Class & 94.89\% \\
    \bottomrule
 \end{tabular}
 \label{Tab1}
\end{table*}   

\subsection{ML Model and Accuracy of Attack Patterns Analysis}
\begin{figure}[h]
    \centering
    \includegraphics[scale=0.52]{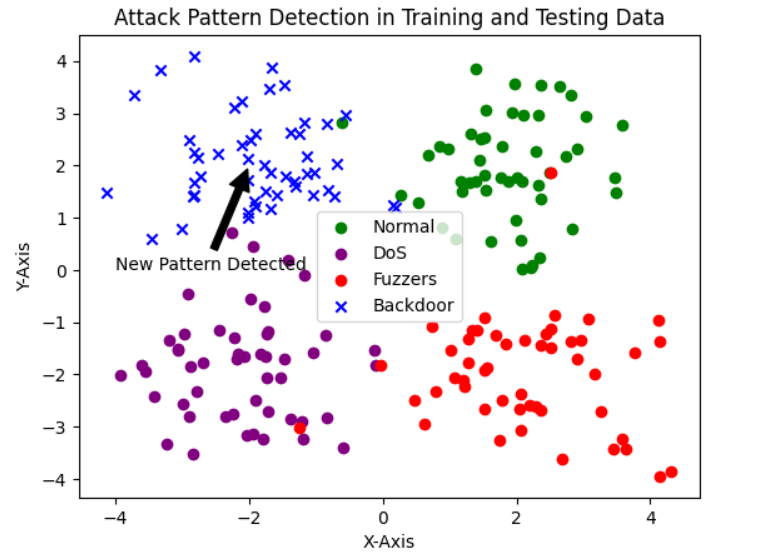}
   \caption{\textcolor{black}{Attack Pattern Detection}}
    \label{fig:f6}
\end{figure}

\begin{figure}[h]
    \centering
    \includegraphics[scale=0.80]{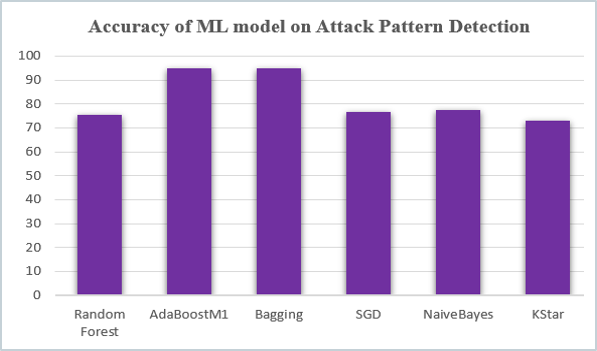}
   \caption{\textcolor{black}{Accuracy Matrics of ML Model}}
    \label{fig:f7}
\end{figure}

As we know zero-day attack is unrecognizable because it has no known pattern based on which we can perform a machine-learning approach. To overcome this issue we developed an XAI-based methodology, that can find out any unknown/newly generated pattern, which is not possible in any regular approach. Through the XAI to get the zero-day attack pattern we generate the SHAP value as given in Fig. \ref{fig:f4}. The attributes/features that we get from the SHAP values are passed to several ML techniques which identify/detect the pattern that is newly encountered. In our dataset there are several attack categories, from there in the training phase we pass three attack categories called Normal, Dos, and Fuzzers \cite{rahman2023federated}. Among them Dos, and Fuzzers are anomalies, and Normal is not anomaly type at all. In the testing phase, we passed a new category of attack called  Blackdoor which is detected by several ML techniques as shown in the Fig. \ref{fig:f6}. Among them, AdaBoostM1 shows the highest accuracy in the case of attack pattern detection as shown in Fig. \ref{fig:f7}.

\subsection{Accuracy and Time of Anomaly Analysis}
\begin{figure}[!ht]
    \centering
    \includegraphics[scale=0.20]{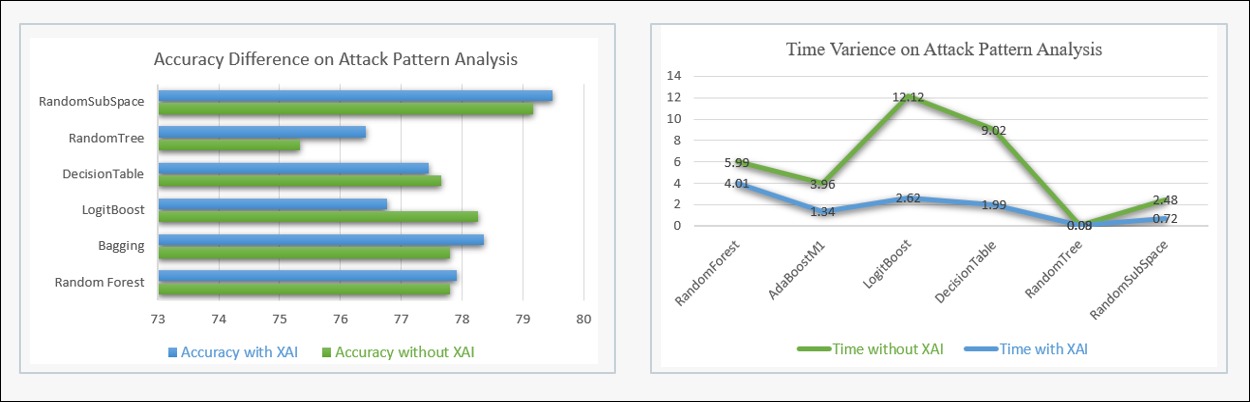}
   \caption{\textcolor{black}{Accuracy difference and Time efficiency}}
    \label{fig:f8}
\end{figure}

As discussed earlier, our proposed method can detect anomalies as a conventional IDPS system in the intermediate layer, yet in a more efficient manner with reduced computational time and improved accuracy. In the case of accuracy from Fig.\ref{fig:f8}, we assert that without XAI, all of the ML models perform less accurately than with XAI, and similar cases also happen for the time factor. In the case of time, after applying XAI for some ML models like LogitBoost and DecisionTable, time drastically fell, which indicates a significant reduction in computational time. Thus, we can say two objectives, attack pattern detection and anomaly detection are performed correctly and in an efficient manner.

\section{Conclusion}
Zero-day is the most vulnerable attack in the cyber security system. It can not be detected by a regular system, as its pattern is unknown and unrecognizable. To address this issue in the smart community in combination with interoperability, we introduce an intermediate layer that detects the unknown pattern of zero-day by XAI and supplies it to the IDPS system, which detects the anomaly. Furthermore, this intermediate layer works as double-layer anomaly detection, where regular attacks are detected in this layer in a more efficient way in the case of time and accuracy, and zero-day is detected in the final layer as XAI detects the pattern and passes it to the IDPS system. In the case of zero-day pattern detection, AdaBoostM1 performed the highest accuracy, whereas in anomaly detection, RandomSubspace triggered the highest accuracy for both with and without XAI. LogitBoost and DecisionTable take the lowest computational time after applying XAI in case of anomaly detection. The objectives that we mentioned before have been successfully completed using our proposed methodology. It can detect any kind of attack, including the most vulnerable zero-day. In the future, we will work on zero-day attack prevention with a larger dataset.

\ifCLASSOPTIONcaptionsoff
  \newpage
\fi

\bibliographystyle{IEEEtran}

\bibliography{sample}

\begin{thebibliography}{10}
\providecommand{\url}[1]{#1}
\csname url@samestyle\endcsname
\providecommand{\newblock}{\relax}
\providecommand{\bibinfo}[2]{#2}
\providecommand{\BIBentrySTDinterwordspacing}{\spaceskip=0pt\relax}
\providecommand{\BIBentryALTinterwordstretchfactor}{4}
\providecommand{\BIBentryALTinterwordspacing}{\spaceskip=\fontdimen2\font plus
\BIBentryALTinterwordstretchfactor\fontdimen3\font minus \fontdimen4\font\relax}
\providecommand{\BIBforeignlanguage}[2]{{%
\expandafter\ifx\csname l@#1\endcsname\relax
\typeout{** WARNING: IEEEtran.bst: No hyphenation pattern has been}%
\typeout{** loaded for the language `#1'. Using the pattern for}%
\typeout{** the default language instead.}%
\else
\language=\csname l@#1\endcsname
\fi
#2}}
\providecommand{\BIBdecl}{\relax}
\BIBdecl

\bibitem{kumar2024detection}
A.~Kumar and B.~K. Chaurasia, ``Detection of sars-cov-2 virus using lightweight convolutional neural networks,'' \emph{Wireless Personal Communications}, vol. 135, no.~2, pp. 941--965, 2024.

\bibitem{Rahman2020}
A.~Rahman, U.~Sara, D.~Kundu, S.~Islam, M.~J. Islam, M.~Hasan, Z.~Rahman, and M.~K. Nasir, ``Distb-sdoindustry: Enhancing security in industry 4.0 services based on distributed blockchain through software defined networking-iot enabled architecture,'' \emph{International Journal of Advanced Computer Science and Applications}, vol.~11, no.~9, 2020.

\bibitem{ahmad2023zero}
R.~Ahmad, I.~Alsmadi, W.~Alhamdani, and L.~Tawalbeh, ``Zero-day attack detection: a systematic literature review,'' \emph{Artificial Intelligence Review}, vol.~56, no.~10, pp. 10\,733--10\,811, 2023.

\bibitem{rizzardi2024nero}
A.~Rizzardi, S.~Sicari, A.~C. Porisini \emph{et~al.}, ``Nero: Neural algorithmic reasoning for zero-day attack detection in the iot: A hybrid approach,'' \emph{Computers \& Security}, p. 103898, 2024.

\bibitem{rahman2021study}
A.~Rahman, M.~Rahman, D.~Kundu, M.~R. Karim, S.~S. Band, and M.~Sookhak, ``Study on iot for sars-cov-2 with healthcare:present and future perspective,'' \emph{Mathematical Biosciences and Engineering}, vol.~18, no.~6, pp. 9697--9726, 2021.

\bibitem{rahman2024machine}
A.~Rahman, T.~Debnath, D.~Kundu, M.~S.~I. Khan, A.~A. Aishi, S.~Sazzad, M.~Sayduzzaman, and S.~S. Band, ``Machine learning and deep learning-based approach in smart healthcare: Recent advances, applications, challenges and opportunities,'' \emph{AIMS Public Health}, vol.~11, no.~1, pp. 58--109, 2024.

\bibitem{rahman2024internet}
A.~Rahman, M.~A.~H. Wadud, M.~J. Islam, D.~Kundu, T.~A.-U.-H. Bhuiyan, G.~Muhammad, and Z.~Ali, ``Internet of medical things and blockchain-enabled patient-centric agent through sdn for remote patient monitoring in 5g network,'' \emph{Scientific Reports}, vol.~14, no.~1, p. 5297, 2024.

\bibitem{rahman2023impacts}
A.~Rahman, J.~Islam, D.~Kundu, R.~Karim, Z.~Rahman, S.~S. Band, M.~Sookhak, P.~Tiwari, and N.~Kumar, ``Impacts of blockchain in software-defined internet of things ecosystem with network function virtualization for smart applications: Present perspectives and future directions,'' \emph{International Journal of Communication Systems}, p. e5429, 2023.

\bibitem{jinhong2024cross}
F.~Jinhong, ``Cross-platform and multi-terminal collaborative software information security strategy,'' in \emph{2024 5th International Conference on Mobile Computing and Sustainable Informatics (ICMCSI)}.\hskip 1em plus 0.5em minus 0.4em\relax IEEE, 2024, pp. 781--787.

\bibitem{9528133}
K.~M. Shayshab~Azad, N.~Hossain, M.~J. Islam, A.~Rahman, and S.~Kabir, ``Preventive determination and avoidance of ddos attack with sdn over the iot networks,'' in \emph{2021 International Conference on Automation, Control and Mechatronics for Industry 4.0 (ACMI)}, 2021, pp. 1--6.

\bibitem{rahman2022integration}
A.~Rahman, A.~Montieri, D.~Kundu, M.~Karim, M.~Islam, S.~Umme, A.~Nascita, A.~Pescap{\'e} \emph{et~al.}, ``On the integration of blockchain and sdn: Overview, applications, and future perspectives,'' \emph{Journal of Network and Systems Management}, vol.~30, no.~4, pp. 1--44, 2022.

\bibitem{32}
R.~Kishore and A.~Chauhan, ``Intrusion detection system a need,'' in \emph{2020 IEEE International Conference for Innovation in Technology (INOCON)}, 2020, pp. 1--7.

\bibitem{31}
K.-A. Tait, J.~S. Khan, F.~Alqahtani, A.~A. Shah, F.~Ali~Khan, M.~U. Rehman, W.~Boulila, and J.~Ahmad, ``Intrusion detection using machine learning techniques: An experimental comparison,'' in \emph{2021 International Congress of Advanced Technology and Engineering (ICOTEN)}, 2021, pp. 1--10.

\bibitem{33}
K.~Saurabh, S.~Sood, P.~A. Kumar, U.~Singh, R.~Vyas, O.~Vyas, and R.~Khondoker, ``Lbdmids: Lstm based deep learning model for intrusion detection systems for iot networks,'' in \emph{2022 IEEE World AI IoT Congress (AIIoT)}, 2022, pp. 753--759.

\bibitem{kumar2021robust}
V.~Kumar and D.~Sinha, ``A robust intelligent zero-day cyber-attack detection technique,'' \emph{Complex \& Intelligent Systems}, vol.~7, no.~5, pp. 2211--2234, 2021.

\bibitem{hindy2020utilising}
H.~Hindy, R.~Atkinson, C.~Tachtatzis, J.-N. Colin, E.~Bayne, and X.~Bellekens, ``Utilising deep learning techniques for effective zero-day attack detection,'' \emph{Electronics}, vol.~9, no.~10, p. 1684, 2020.

\bibitem{34}
M.~Macas and C.~Wu, ``Review: Deep learning methods for cybersecurity and intrusion detection systems,'' in \emph{2020 IEEE Latin-American Conference on Communications (LATINCOM)}, 2020, pp. 1--6.

\bibitem{zhang2021novel}
C.~Zhang, Y.~Chen, Y.~Meng, F.~Ruan, R.~Chen, Y.~Li, and Y.~Yang, ``A novel framework design of network intrusion detection based on machine learning techniques,'' \emph{Security and Communication Networks}, vol. 2021, no.~1, p. 6610675, 2021.

\bibitem{martins2022host}
I.~Martins, J.~S. Resende, P.~R. Sousa, S.~Silva, L.~Antunes, and J.~Gama, ``Host-based ids: A review and open issues of an anomaly detection system in iot,'' \emph{Future Generation Computer Systems}, vol. 133, pp. 95--113, 2022.

\bibitem{salimarticlesfederated}
M.~M. Salim, Y.~Sangthong, X.~Deng, and J.~H. Park, ``Articlesfederated learning-enabled zero-day ddos attack detection scheme in healthcare 4.0,'' vol.~14, 2024.

\bibitem{zahoora2022ransomware}
U.~Zahoora, A.~Khan, M.~Rajarajan, S.~H. Khan, M.~Asam, and T.~Jamal, ``Ransomware detection using deep learning based unsupervised feature extraction and a cost sensitive pareto ensemble classifier,'' \emph{Scientific Reports}, vol.~12, no.~1, p. 15647, 2022.

\bibitem{efe2022comparison}
A.~Efe and {\.I}.~N. Abac{\i}, ``Comparison of the host based intrusion detection systems and network based intrusion detection systems,'' pp. 23--32, 2022.

\bibitem{35}
J.~Nie, P.~Ma, B.~Wang, and Y.~Su, ``A covert network attack detection method based on lstm,'' in \emph{2020 IEEE 5th Information Technology and Mechatronics Engineering Conference (ITOEC)}, 2020, pp. 1690--1693.

\bibitem{talukder2023dependable}
M.~A. Talukder, K.~F. Hasan, M.~M. Islam, M.~A. Uddin, A.~Akhter, M.~A. Yousuf, F.~Alharbi, and M.~A. Moni, ``A dependable hybrid machine learning model for network intrusion detection,'' \emph{Journal of Information Security and Applications}, vol.~72, p. 103405, 2023.

\bibitem{36}
C.~Redino, D.~Nandakumar, R.~Schiller, K.~Choi, A.~Rahman, E.~Bowen, A.~Shaha, J.~Nehila, and M.~Weeks, ``Zero day threat detection using graph and flow based security telemetry,'' in \emph{2022 International Conference on Computing, Communication, and Intelligent Systems (ICCCIS)}, 2022, pp. 655--662.

\bibitem{rahman2020distblockbuilding}
A.~Rahman, M.~K. Nasir, Z.~Rahman, A.~Mosavi, S.~Shahab, and B.~Minaei-Bidgoli, ``Distblockbuilding: A distributed blockchain-based sdn-iot network for smart building management,'' \emph{IEEE Access}, vol.~8, pp. 140\,008--140\,018, 2020.

\bibitem{rahman2020distb}
A.~Rahman, M.~J. Islam, Z.~Rahman, M.~M. Reza, A.~Anwar, M.~P. Mahmud, M.~K. Nasir, and R.~M. Noor, ``Distb-condo: Distributed blockchain-based iot-sdn model for smart condominium,'' \emph{IEEE Access}, vol.~8, pp. 209\,594--209\,609, 2020.

\bibitem{10103295}
M.~Faisal, H.~Siddiqua, M.~J. Islam, and A.~Rahman, ``An sdn-based secure model for iot network in smart building,'' in \emph{2022 4th International Conference on Sustainable Technologies for Industry 4.0 (STI)}, 2022, pp. 1--6.

\bibitem{rahman2021smartblock}
A.~Rahman, M.~J. Islam, A.~Montieri, M.~K. Nasir, M.~M. Reza, S.~S. Band, A.~Pescape, M.~Hasan, M.~Sookhak, and A.~Mosavi, ``Smartblock-sdn: An optimized blockchain-sdn framework for resource management in iot,'' \emph{IEEE Access}, vol.~9, pp. 28\,361--28\,376, 2021.

\bibitem{9499121}
M.~J. Islam, A.~Rahman, S.~Kabir, M.~R. Karim, U.~K. Acharjee, M.~K. Nasir, S.~S. Band, M.~Sookhak, and S.~Wu, ``Blockchain-sdn-based energy-aware and distributed secure architecture for iot in smart cities,'' \emph{IEEE Internet of Things Journal}, vol.~9, no.~5, pp. 3850--3864, 2022.

\bibitem{rahman2023icn}
A.~Rahman, K.~Hasan, D.~Kundu, M.~J. Islam, T.~Debnath, S.~S. Band, and N.~Kumar, ``On the icn-iot with federated learning integration of communication: Concepts, security-privacy issues, applications, and future perspectives,'' \emph{Future Generation Computer Systems}, vol. 138, pp. 61--88, 2023.

\bibitem{edvardsson1999survey}
J.~Edvardsson, ``A survey on automatic test data generation,'' in \emph{Proceedings of the 2nd Conference on Computer Science and Engineering}, 1999, pp. 21--28.

\bibitem{rahman2022sdn}
A.~Rahman, C.~Chakraborty, A.~Anwar, M.~Karim, M.~Islam, D.~Kundu, Z.~Rahman, S.~S. Band \emph{et~al.}, ``Sdn--iot empowered intelligent framework for industry 4.0 applications during covid-19 pandemic,'' \emph{Cluster Computing}, vol.~25, no.~4, pp. 2351--2368, 2022.

\bibitem{moustafa2016evaluation}
N.~Moustafa and J.~Slay, ``The evaluation of network anomaly detection systems: Statistical analysis of the unsw-nb15 data set and the comparison with the kdd99 data set,'' \emph{Information Security Journal: A Global Perspective}, vol.~25, no. 1-3, pp. 18--31, 2016.

\bibitem{udoy20234sqr}
A.~I. Udoy, M.~A. Rahaman, M.~J. Islam, A.~Rahman, Z.~Ali, and G.~Muhammad, ``4sqr-code: A 4-state qr code generation model for increasing data storing capacity in the digital twin framework,'' \emph{Journal of Advanced Research}, 2023.

\bibitem{ahmed2023image}
M.~T. Ahmed, R.~Islam, M.~A. Rahman, M.~J. Islam, A.~Rahman, and S.~Kabir, ``An image-based digital forensic investigation framework for crime analysis,'' in \emph{2023 International Conference on Next-Generation Computing, IoT and Machine Learning (NCIM)}.\hskip 1em plus 0.5em minus 0.4em\relax IEEE, 2023, pp. 1--6.

\bibitem{demertzis2019next}
K.~Demertzis, N.~Tziritas, P.~Kikiras, S.~L. Sanchez, and L.~Iliadis, ``The next generation cognitive security operations center: adaptive analytic lambda architecture for efficient defense against adversarial attacks,'' \emph{Big Data and Cognitive Computing}, vol.~3, no.~1, p.~6, 2019.

\bibitem{soltani2023adaptable}
M.~Soltani, B.~Ousat, M.~J. Siavoshani, and A.~H. Jahangir, ``An adaptable deep learning-based intrusion detection system to zero-day attacks,'' \emph{Journal of Information Security and Applications}, vol.~76, p. 103516, 2023.

\bibitem{rahman2023towards}
A.~Rahman, M.~J. Islam, S.~S. Band, G.~Muhammad, K.~Hasan, and P.~Tiwari, ``Towards a blockchain-sdn-based secure architecture for cloud computing in smart industrial iot,'' \emph{Digital Communications and Networks}, vol.~9, no.~2, pp. 411--421, 2023.

\bibitem{dhanabal2015study}
L.~Dhanabal and S.~Shantharajah, ``A study on nsl-kdd dataset for intrusion detection system based on classification algorithms,'' \emph{International journal of advanced research in computer and communication engineering}, vol.~4, no.~6, pp. 446--452, 2015.

\bibitem{khan2022multinet}
S.~I. Khan, A.~Shahrior, R.~Karim, M.~Hasan, and A.~Rahman, ``Multinet: A deep neural network approach for detecting breast cancer through multi-scale feature fusion,'' \emph{Journal of King Saud University-Computer and Information Sciences}, vol.~34, no.~8, pp. 6217--6228, 2022.

\bibitem{kundu2024federated}
D.~Kundu, M.~M. Rahman, A.~Rahman, D.~Das, U.~R. Siddiqi, M.~G.~R. Alam, S.~K. Dey, G.~Muhammad, and Z.~Ali, ``Federated deep learning for monkeypox disease detection on gan-augmented dataset,'' \emph{IEEE Access}, 2024.

\bibitem{Islam2021}
S.~Islam, U.~Sara, A.~Kawsar, A.~Rahman, D.~Kundu, D.~D. Dipta, A.~R. Karim, and M.~Hasan, ``Sgbba: An efficient method for prediction system in machine learning using imbalance dataset,'' \emph{International Journal of Advanced Computer Science and Applications}, vol.~12, no.~3, 2021.

\bibitem{Debnath2022}
\BIBentryALTinterwordspacing
T.~Debnath, M.~M. Reza, A.~Rahman, A.~Beheshti, S.~S. Band, and H.~Alinejad-Rokny, ``{Four-layer ConvNet to facial emotion recognition with minimal epochs and the significance of data diversity},'' \emph{Scientific Reports}, vol.~12, no.~1, p. 6991, dec 2022. [Online]. Available: \url{https://www.nature.com/articles/s41598-022-11173-0}
\BIBentrySTDinterwordspacing

\bibitem{moustafa2021new}
N.~Moustafa, ``A new distributed architecture for evaluating ai-based security systems at the edge: Network ton\_iot datasets,'' \emph{Sustainable Cities and Society}, vol.~72, p. 102994, 2021.

\bibitem{rahman2024blocksd}
A.~Rahman, M.~S.~I. Khan, A.~Montieri, M.~J. Islam, M.~R. Karim, M.~Hasan, D.~Kundu, M.~K. Nasir, and A.~Pescap{\`e}, ``Blocksd-5gnet: Enhancing security of 5g network through blockchain-sdn with ml-based bandwidth prediction,'' \emph{Transactions on Emerging Telecommunications Technologies}, vol.~35, no.~4, p. e4965, 2024.

\bibitem{sarhan2023zero}
M.~Sarhan, S.~Layeghy, M.~Gallagher, and M.~Portmann, ``From zero-shot machine learning to zero-day attack detection,'' \emph{International Journal of Information Security}, vol.~22, no.~4, pp. 947--959, 2023.

\bibitem{koroniotis2020new}
N.~Koroniotis, N.~Moustafa, and E.~Sitnikova, ``A new network forensic framework based on deep learning for internet of things networks: A particle deep framework,'' \emph{Future Generation Computer Systems}, vol. 110, pp. 91--106, 2020.

\bibitem{ashraf2020novel}
J.~Ashraf, A.~D. Bakhshi, N.~Moustafa, H.~Khurshid, A.~Javed, and A.~Beheshti, ``Novel deep learning-enabled lstm autoencoder architecture for discovering anomalous events from intelligent transportation systems,'' \emph{IEEE Transactions on Intelligent Transportation Systems}, vol.~22, no.~7, pp. 4507--4518, 2020.

\bibitem{rahman2023federated}
A.~Rahman, M.~S. Hossain, G.~Muhammad, D.~Kundu, T.~Debnath, M.~Rahman, M.~S.~I. Khan, P.~Tiwari, and S.~S. Band, ``Federated learning-based ai approaches in smart healthcare: concepts, taxonomies, challenges and open issues,'' \emph{Cluster computing}, vol.~26, no.~4, pp. 2271--2311, 2023.

\end{thebibliography}

\end{document}